\documentstyle[epsfig]{aipproc}

\begin{document}

\title{Brane World    Astronomy}

\author{Craig J. Hogan}

\address{ Astronomy and Physics Departments, University of Washington,
Seattle, WA 98195-1580, USA}

\maketitle

\begin{abstract} Unified theories suggest that space  is intrinsically
10  dimensional, even though everyday phenomena seem to take place in
only 3  large dimensions. 
 In ``Brane World'' models,    matter and radiation are localized to a ``brane''
which has a thickness less than $\approx ({\rm TeV})^{-1}$ in all but the usual three
  dimensions, while gravity   propagates in   additional dimensions, some 
of  which may   extend as far as submillimeter scales. A brief review is presented of  
some of these models and their astrophysical phenomenology. 
One distinctive possibility  is a gravitational wave background originating in   the
mesoscopic early universe, at temperatures above about 1 TeV and  on scales smaller than a
millimeter,   during the formation of   our 3-dimensional brane within a 
10-dimensional space.
\end{abstract}
\section{Unified Theory}
The Standard Model of strong, weak and electromagnetic  interactions includes all the
forms of mass-energy so far observed  in nature, other than gravity.
It is  based on a relativistic quantum field theory of interacting
fermion  and boson fields,
with forces arising from
 Yang-Mills vector gauge fields,  propagating in a
3+1-dimensional spacetime. Gravity is formulated in a completely different way, using General
Relativity, as a  classical theory of dynamical spacetime itself: ``Spacetime tells mass-energy
how to move, and mass-energy tells spacetime how to curve.'' 

Even though there is no direct inconsistency or disagreement
of these theories with experiment or with each other,
there is
widespread dissatisfaction with the inelegance of this dualistic  situation.  It is
suspected by those who believe in the unity of the natural world that there might be a
single unified theory, derivable from simple principles of symmetry,
 which will appear
in an appropriate mathematical limit  as  the Standard Model fields propagating in a
General Relativistic spacetime.

Major steps   have been taken recently
in the construction  of a unified theory;
one can cite several triumphs of  a theoretical nature, such as the 
ability to count precisely the quantum mechanical degrees of freedom
of spacetime itself  in
certain special black hole spacetimes \cite{horowitz}.   We may now also have   
observed \cite{riess,perlmutter,suntzeff}
 the first real-world  phenomenon which specifically
calls upon new quantum-gravity unification physics  outside of  ``Standard Model Plus GR'': the
Cosmological Constant  (or ``Dark Energy''). Thus there is real hope that the Theory of
Everything may become a real, testable physical theory. 
However it is not clear how to complete
the most important step, connecting the
fundamental theory to the real world which is so well
described by ``Standard Model Plus GR.''
The current  best candidate for a Theory of Everything, ``M theory'',
 is formulated in ten spatial dimensions instead of three,
and   has no direct, distinctive connections with any real-world experiment.

Recent developments in M theory suggest that there may be an  intermediate
level of structure associated with dimensional reduction, which has spawned
a wide variety of proposed designs for new ``Brane World'' models
\cite{arkanihierarchy,antoniadas,arkaniphenom,randallsundrum1,randallsundrum2,lykkenrandall,karchrandall,sundrum}:
In these   models,   the fields of the Standard Model are confined to an
approximately  
 three-dimensional wall or
``brane'' imbedded in an extended ten-dimensional space or ``bulk'', which is
described  by adding extra dimensions to  General Relativity.
The brane has a thickness smaller than the TeV scale
of current particle experiments, while the extra dimensions 
of the bulk can be as large  as the $10^{-2}$eV scale of current gravity 
experiments.   This paper is a  
brief   overview  of
 brane worlds and some new effects they might produce in  astrophysics. 

Brane world models aim to
 short-cut
the connection between fundamental theory and phenomena. They  introduce a kind of
``effective theory'' as a  conceptual bridge--- a parametrized  model broadly motivated
by structures in the fundamental theory, which can be used to calculate  new phenomena
at low energy. Although it is not clear that this strategy will work in the 
long run, it has certainly broken a logjam in thinking and has spawned many
intriguing new  theoretical predictions and experimental tests.

\section{ extra dimensions}

Direct detailed  data from accelerators  confirm the 3+1-dimensional behavior of
Standard Model quantum fields   directly to the current experimental energies
of about 100 GeV, and with some modest extrapolation to about 1 TeV.  That
is, any effects of a fourth spatial dimension had better not appear
in particle interactions unless it is on a length scale much smaller than
$(1\ {\rm TeV})^{-1}$. 

The ideas \cite{wilczek,weinberg} of the logarithmic  running  of couplings  in
standard supersymmetric  grand unification (SUSY GUTs)  suggest an extension 
to much higher energies. Renormalization group calculations allow 
extrapolation of the observed strong, weak and electromagnetic couplings
as a function of energy; the three
curves intersect at the supersymmetric grand unification scale, around $10^{16}$ GeV. This
nontrivial intersection, which has been confirmed by increasingly precise accelerator
experiments, is  often cited as evidence for the unification scheme. In SUSY GUTs, the Standard
Model structure including 3+1-D field theory is preserved up to this much higher energy
scale.

The limits on the dimensionality of gravity are much weaker. Gauss' law
tells us that the gravitational force falls off as $r^{1-N_s}$ where $N_s$ is
the total number of spatial dimensions. Experiments\cite{adelberger} (motivated in part
by  brane world models) now confirm Newton's
inverse square law (and hence $N_s=3$) from astronomical scales down to hair's-width distances
of about 250$\mu$m. This  submillimeter scale 
is is however still vastly larger than those probed by the Standard
Model fields; 0.3 mm corresponds to an energy of about 0.003 eV.

The ``Cosmological Constant Problem'' may be related to effects of extra dimensions.
One way to state this problem (there are many) begins with the observation that the zero point
fluctuations  of Standard Model quantum fields on a given scale $E$, if they couple to gravity,
correspond to a gravitating vacuum energy density of magnitude $E^4$.  The observed
vacuum energy (or Dark Energy) density is about equal to the critical density
of the universe, which is about $(10^{-2} {\rm eV})^4$. But the success of the
Standard Model requires the presence of the field fluctuations  
at least up to TeV scales. (Above that energy, it is possible for fermion and boson 
contributions to cancel exactly due to supersymmetry. Below that energy,
we know that the system is not supersymmetric in today's vacuum). As presently
formulated, the theory requires an offset of the zero energy level of the vacuum magically tuned
to a precision of $17\times 4$
orders of magnitude.

If for some reason the coupling of gravity to zero point modes were strongly suppressed
above  $ 10^{-2} {\rm eV}  $, the gravitating energy of the vacuum would come 
out about right. The corresponding length scale, 0.1mm, lies just below the 
current experimental tests for gravitational coupling, but will become accessible with
the next generation experiments.  (Note that the predictions of the brane worlds
with extra dimensions on these scales, which are discussed below, naitvely have the wrong
sign to solve the cosmological constant problem, since 
with $N_s>3$ the gravitational force  increases faster
than
$r^{-2}$ on small scales).

A  coincidence worth mentioning is that the gravitational timescale associated
with matter at an energy density of $\rho\approx  ({\rm TeV})^4$ is $(G\rho)^{-1/2}\approx
1mm/c$. This coincidence is important in cosmology because it means that the
uncertainty associated with the possible geometrical effects of gravity propagating in extra
dimensions start at about the same place as the uncertainties associated with the 
possible new physics beyond the TeV scale. If there is indeed new physics 
at the 0.1mm scale responsible for the cosmological constant, then this also
would explain the coincidence between the age of the universe when the cosmological
constant starts to dominate the mass-energy (that is, about now), and the
typical lifetimes of stars; that is, there may be a derivable reason why
$\rho_{vacuum}\approx M_{Planck}^4 (M_{proton}/M_{Planck})^6$.

Aside from the cosmological constant, there is the 
corresponding  ÒHierarchy ProblemÓ in particle physics itself: if the fundamental scale
is $10^{19}$ GeV, how is the ``light'' TeV scale preserved in all orders of 
all interactions?  (Related to this: what is the
origin of ÒLarge NumbersÓ of astrophysics, which derive from the large
ratio $m_{Planck}/m_{proton}$?) The traditional approach is to invoke 
supersymmetry above the TeV scale to preserve the hierarchy, and to explain
the large numbers as due to the logarithmically running couplings. In some of the 
new schemes however, the Planck scale is a kind of illusion; physics is fundamentally different
above the $\approx 10$ TeV scale, which is the only fundamental scale in the theory.
The large numbers in these schemes arise from taking modest numbers to large 
powers; the gravitational force in ten dimensional space falls off as
$r^{-9}$!

\section{ extra dimensions in unified theory}

The candidate Theory of Everything, sometimes called ÒM theoryÓ 
(or supersymmetric  superstring theory, or matrix theory, etc., depending
on the limit and the context),
consistently includes both quantum mechanics
and general relativity, and possibly includes the  Standard Model.
It provides a framework for computing statistically the   entropy
of certain black holes from first principles.
 A hallmark of the theory is a  formal melding and blurring of the 
distinction between  string (and particle) degrees of freedom,  and
geometrical degrees of freedom.
Powerful  dualities are exploited to show the equivalence of different
formulations and  between
large and small scales and strong and weak coupling limits. 
In spite of the fluid   character of the ideas, one central property seems to
become more firmly established with time: the theory exists only
in 10 fundamental spatial dimensions.

The central idea for dealing with  the   7 ``extra'' dimensions  is 
``compactification'':
we are unaware of the extra dimensions because they are much smaller than 
the three normal large space dimensions. For example,  a 
three dimensional tube   can appear two  dimensional if its walls are thin
enough, and even one dimensional if it is long and thin enough. As the above remarks
indicate, the size of the extra dimensions for Standard Model field propagation
must be smaller than  $({\rm TeV})^{-1}$, and smaller than $(10^{16}{\rm GeV})^{-1}$
if SUSY GUT ideas are right.
 
The idea  of compactification was investigated by
Kaluza and Klein in the 1920's as a way to unify gravity and electromagnetism.
They showed the geometrodynamics of an additional very  small dimension could appear at low
energies as an electromagnetic field tensor.  Extra dimensions
lead to new predicted
degrees of freedom in fields--- for example an intrinsically massless
 field  creates a  ``Kaluza- Klein tower'' of new, effectively massive
excitations corresponding to harmonics of states propagating around the short new directions. 
Altough traditionally the extra dimensions and new effects associated with
M-theory have generally been assumed to happen close to the
Planck scale, this is not neccesarily the case; the most notable result of brane world
models is that the extra dimensions may be very large, possibly even infinite in extent.

M-theory is known to contain   structures that 
  offer suggestive clues to compactification.
Features called ÒbranesÓ appear which
have  lower dimensionality than the whole space. They form sites where
the fundamental objects, one dimensional strings, can terminate, suggesting that in a low
energy theory the gauge interactions may be confined to a lower dimensional surface
embedded in the ten dimensional space.
Branes are often thought of as 
classical, defect- or soliton-like structures resembling cosmic domain walls, with a surface
tension and an internal vacuum energy larger than that
in the surrounding higher dimensional background space.
(Since the latter can be negative, the background space can be a higher-dimensional
 Anti-de
Sitter solution even though our universe has a positive energy density).
 New types of
excitations, corresponding to degrees of freedom such as displacement of the brane in the higher
dimensions, correspond to propagating modes and  new types of particles that might be  observed. 

One of the  most spectacular discoveries in   unification theory   is Maldacena's
AdS5/CFT correspondence: 
$N=4$ supergravity in extended  5D Anti-deSitter space
is exactly equivalent  to a conformal field theory on its 4D
boundary, which can be regarded as just ordinary Minkowski space. Here is a concrete example of a
quantum gravity theory with all the richness associated with fields in five dimensions,
all the details of which map onto the behavior of a conformal field theory in 
the standard 4D spacetime\cite{witten}. 

A  related   earlier idea called 
``holography''   was inspired by the thermodynamics and information content
 of black holes.  
The conjecture is that all 3D fields are actually
encoded by some theory acting on a 2D surface. We  know that black hole
entropy is given by a constant times the surface area of its (two-dimensional)
event horizon. This means that a   finite (indeed countable) amount of 
 data on a two-dimensional surface (roughly, a few bytes per Planck area)
must suffice to specify everything going on in the three-dimensional volume of space within it. 
The holography conjecture is that this applies to the whole universe--- that 
three-dimensional space in some sense is an illusion, that the actual behavior
is in some more fundamental sense  two dimensional.

\section{brane worlds}

Brane worlds start with the
idea that the familiar  
 Standard Model fields are confined to
  a   wall or ``brane''. This structure has three large  dimensions
but  a thickness  
TeV$^{-1}$ or smaller in the other dimensions.  Gravity on the other hand
can  propagate much farther into one or
more   larger  dimensions (called the ``bulk'').  
The brane can be thought of as a stable classical defect embedded in 
a highly symmetric  space of more dimensions, usually Anti-deSitter.
 Within this framework there are many options.

\begin{figure}[htbp] 
\centerline{\epsfig{file=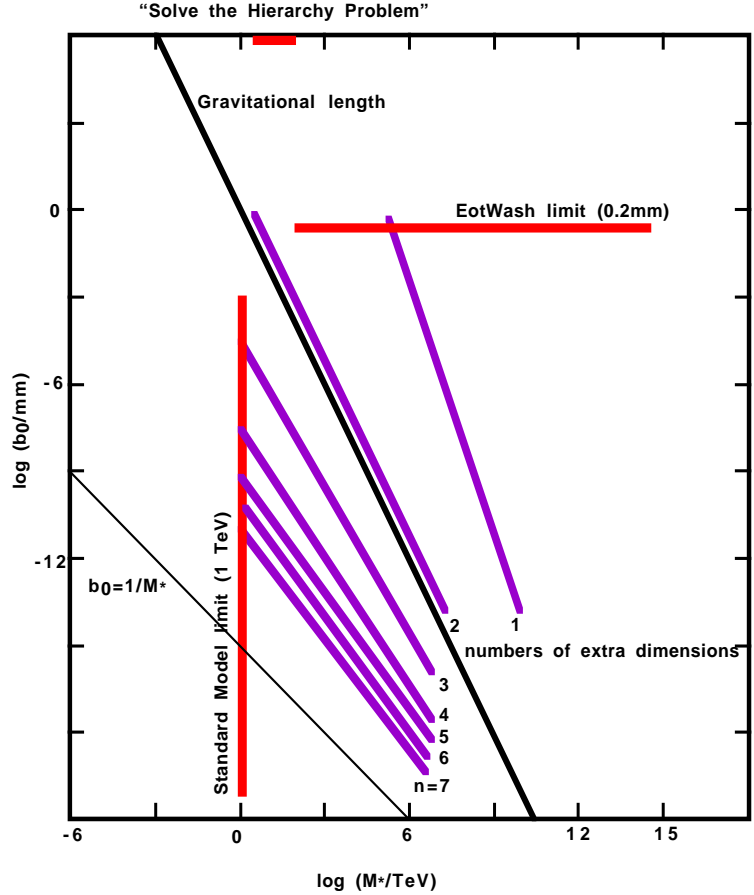,height=5in,width=4in}}
\vspace{0.25in}
\caption{ \label{fig: branescales} Summary of parameters for simple
brane world models. It is assumed that there is a single unification scale
$M_*$ and that there are $n$ extra dimensions of equal size $b$. The series
of lines labeled $n=1$ to 7 correspond to   viable models with the
right gravitational coupling at low energies. 
The ``Gravitational length'' line, degenerate with $n=2$, denotes the 
Schwarzschild radius of a black hole with mean density $M_*^4$.
The thresholds of direct current particle and gravity experiments are shown.
Models with only one fundamental scale, which may ``solve the hierarchy problem'',
lie not far beyond the reach of current accelerator constraints. }
\end{figure}

 In some models, one or two extra dimensions can be of surprisingly large
size\cite{arkanihierarchy,antoniadas,arkaniphenom},  as they are only constrained by
the direct 
experimental gravitational probes of the order of  a 
few hundred microns. Most of  the ``large'' extra dimensions however
must be much smaller than this. 
Elaborations of brane worlds have been explored; for example, some have 
multiple branes which interact gravitationally. In others, there are different branes
 for different Standard Model fermion fields, with 
bosons allowed to travel in the bulk between them.

In one interesting  class of brane-world models (``nonfactorizable geometries'',
\cite{randallsundrum1,randallsundrum2,lykkenrandall,karchrandall}),
the extra dimensions can be even larger, but the larger embedding space is 
highly curved,      
  which traps gravitons in a bound state close to a  brane.
(Such geometries are said to have a ``warp factor'' (!)). 
The curvature radius of  higher-dimensional  (e.g. Anti-deSitter) space  
 is again on a mesoscopic scale, which may be    as large as
$\approx 0.2$ mm. Macroscopic black holes can be pictured as thin 
pancakes stuck to the brane, with only three large dimensions. The AdS space is Poincar\'e
invariant and is itself a stable solution, so the setup is dynamically self-consistent,
the kind of   structure which might develop naturally from a defect
in fields in higher dimensions.

The apparent
(usual) Planck mass in 3+1D, $M_{Planck}$, is related to the true 
fundamental scale $M_*$ by
$M_{Planck}^2\approx M_*^2 (M_*^nV_n)$
where $V_n$ is the volume of $n$ extra   dimensions (larger than $M_*^{-1}$)
in which gravity
propagates. Thus if there is one extra   dimension much larger than the others, the mm limits
from gravity experiments require a unification scale
$M_*=10^6$TeV or larger. If there are  two extra large dimensions the mm limits
give $M_*$ close to the TeV scale. If we try to solve the hierarchy problem
with a single  $M_*$ not too far above the TeV scale, this can be accomplished for $n\ge 2$
by choosing suitable extra dimension sizes; for 
7 equally  large extra dimensions, we might have gravity propagating
in ten dimensions, seven of which have size $b\approx 10^{-10}$ mm.
The range of options for a simple model with $n$ extra
bulk dimensions of the same size $b_0$ is illustrated in figure 1.

An interesting result is that the unification implied by the running-together
 of Standard Model coupling constants can still work  in brane-world scenarios, but the
three gauge couplings come together at a much reduced 
energy \cite{Dienes:1998qh,Dienes:1999vg}. With the addition of extra dimensions
{\it for the gauge fields} (as well as gravity), the
renormalization of the fields produces a power-law dependence of coupling on energy  (like
gravity always had), so that they run together in a rather modest range of energy. For example, 
if the brane has a width in a single extra dimension
of  TeV$^{-1}$ then at higher energy the Standard Model
couplings rapidly  converge and   meet in a point at about 20 TeV.  This 
is regarded 
as less elegant than  the parameter-free running-together of SUSY at 
the $10^{16}$ GeV GUT scale but it may be the way nature works.   These schemes thus hold out the
attractive possibility of a unification scheme, even including gravity,
 with just one scale; it
is  even possible that we might find full quantum
gravity effects accessible at the level of the next-generation accelerators.  
The famous ``Desert'' and the Planck scale, which have shaped so much
discussion in the past,  may be mirages.

\section{brane astronomy}

The new fields and  particles of these models might appear  at accelerators
 in various manifestations. Some of these appear as ``normal'' new particle
effects, such as excitation of Nambu-Goldstone modes of brane oscillations which would
show up with the same signatures of missing energy and momentum as a
weakly interacting scalar particle, or radion modes which might appear
with signatures resembling (if not identical with) a Higgs scalar.
Other possibilities include  quirky signatures, such as multiple, evenly-spaced events 
produced as a particle traveling in the bulk punches periodically through the 3-brane.
Null results in laboratory searches for   measurable departures from Newton's inverse square law
at short distances are an important constraint for $n=1 $ or 2.  

As usual, astrophysical environments reach farther into 
parameter space.   New weakly coupled
species in these models are constrained in the same way axions are, using
arguments based on  energy losses from
supernovae and  red giants.  The
``Kaluza-Klein tower'' states can be particularly interesting.
Massive KK modes of the graviton are a generic effect, and their 
cosmological  production
is an important constraint. They are produced thermally in the early universe,
and only avoid causing an overclosure catastrophe in some cases because they can be
very  weakly
coupled to the thermal particles on the brane.  By the same token, 
for the right parameters they are a cold dark matter
candidate. KK ladders of massive sterile
neutrinos are a possible  candidate for warm dark matter, and may
display unusual nonthermal energy distributions induced by species
oscillations.  Brane worlds may bring important new insights into the 
cosmological constant\cite{flanagan} and inflation\cite{maartens}.
It is even possible for gravitational waves to travel faster than light
since they can take a ``short cut'' across the bulk.

A more speculative phenomenon,
which potentially reaches even farther into parameter space,
is the classical production of gravitational waves
in the early universe,  which survive to the present as a nonthermal stochastic
background\cite{maggiore}.  
  Brane-world models suggest new sources of stochastic backgrounds:
their    new geometrical
 degrees of freedom  can be coherently excited by
symmetry breaking in the early universe, leading to   
gravitational radiation today at   
redshifted frequencies appropriate for    new
observatories such as LIGO and LISA \cite{hoganbrane,hoganbrane2}.
 New
extra-dimensional effects    remain important  until the   Hubble length
 $ H^{-1}\approx M_{Planck}/T^{2} $ is comparable to  the
size or curvature radius $b$ of the  extra dimensions\cite{arkaniphenom,csaki,rapidinflation},
or until the temperature falls below the new unification scale, whichever happens last.

 Of particular interest   are two new geometrical degrees of freedom common to 
many of these models: ``radion''
modes controlling the size or curvature of the extra dimensions\cite{csaki,goldbergerwise},
and new Nambu-Goldstone modes corresponding to inhomogeneous  displacements
 of the brane in the extra dimensions\cite{arkaniphenom,sundrum}.
Cosmological symmetry breaking can   create large-amplitude, coherent classical excitations 
on   scales  of order $H^{-1}$    
as the configuration of the extra dimensions and the position of the brane settle into their
present state.

The scalar modes of this distortion   have long ago disappeared since they
are on a very small scale
 (i.e., less than 1 mm times the redshift, or about the size of the solar system today), but the
tensor modes might be observable.
  Extra dimensions with scale
between 10 \AA\  and 1 mm, which enter the  3+1-D  era at cosmic temperatures 
between 1 and 1000 TeV,    produce backgrounds with energy peaked at observed 
frequencies in the LISA band, between  $10^{-1}$ and $10^{-4}$ Hz. The
background  is detectable above instrument and astrophysical foregrounds  if  
initial metric perturbations are excited to a fractional amplitude of $\approx 10^{-3}$
or  more. As shown in Figure 2, brane world models which ``solve the hierarchy problem'' naturally
produce backgrounds in the range of frequencies encompassed by   LISA
for all the viable cases, $n=2 $ to 7. Ground based 
detectors (LIGO, VIRGO, TAMA, GEO),      probing  higher frequencies, reach
extra dimensions down to $10^{-15}$ mm and unification energies up to $10^{13}$ GeV.

\begin{figure}[htbp] 
\centerline{\epsfxsize=4truein \epsfbox{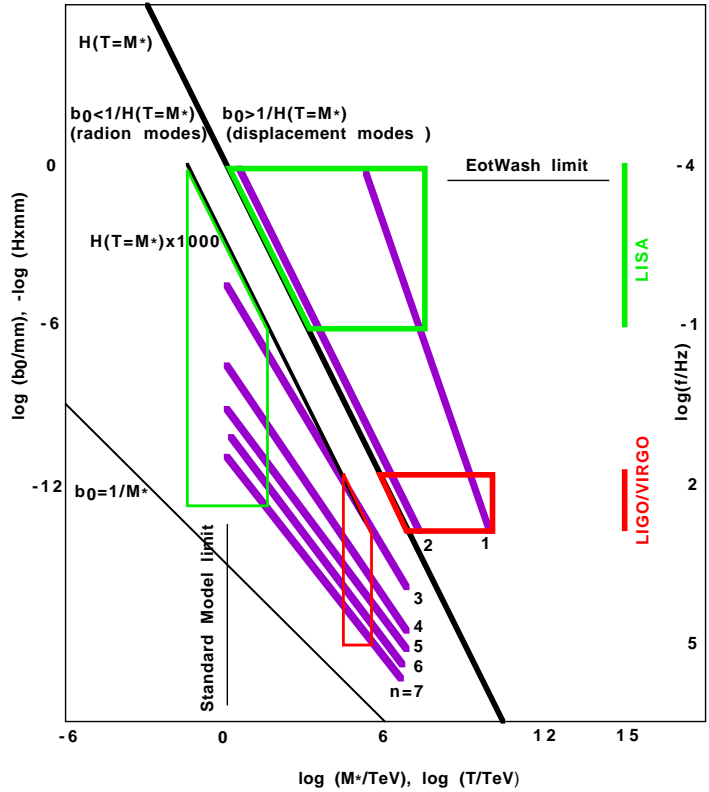}}
\vspace{0.25in}
\caption{ \label{fig: variousn}Summary of the new parameter space of extra dimensions that will be
probed by gravitational-wave interferometers. 
 Boxes indicate the corresponding regions of these
parameters which may give rise to   detectable mesoscopic gravitational radiation backgrounds
in the LISA and LIGO bands.
Heavy-line boxes show the displacement mode parameters, lighter-line
boxes show the radion mode parameters.  These regions extend well beyond those already
constrained by gravitational experiments, direct particle production, or other astrophysical 
constraints. Theories which ``solve the hierarchy problem'' have $M_*$
close to the Standard Model limit, and all of the viable ones ($2\le n\le 7$)
could possibly produce an observable background of one type or the
other in the LISA band.}
\end{figure}

Thus it is  possible that
  gravitational wave astrophysics might  ``see'' outside
of the four dimensions of ordinary spacetime, and   trace the details of how the
three spatial dimensions settled into their present shape, in   brane worlds 
 that cannot be tested by any other known technique (see Figure  2). 
The background spectrum also contains information about 
 a  regime of cosmic history not preserved by
any other relic (Figure 3).

I am grateful for useful conversations with E. Adelberger, D. Kaplan,
 A. Nelson, C. Stubbs,  and S. Weinberg.

\begin{figure}[htbp] 
\centerline{\epsfig{file=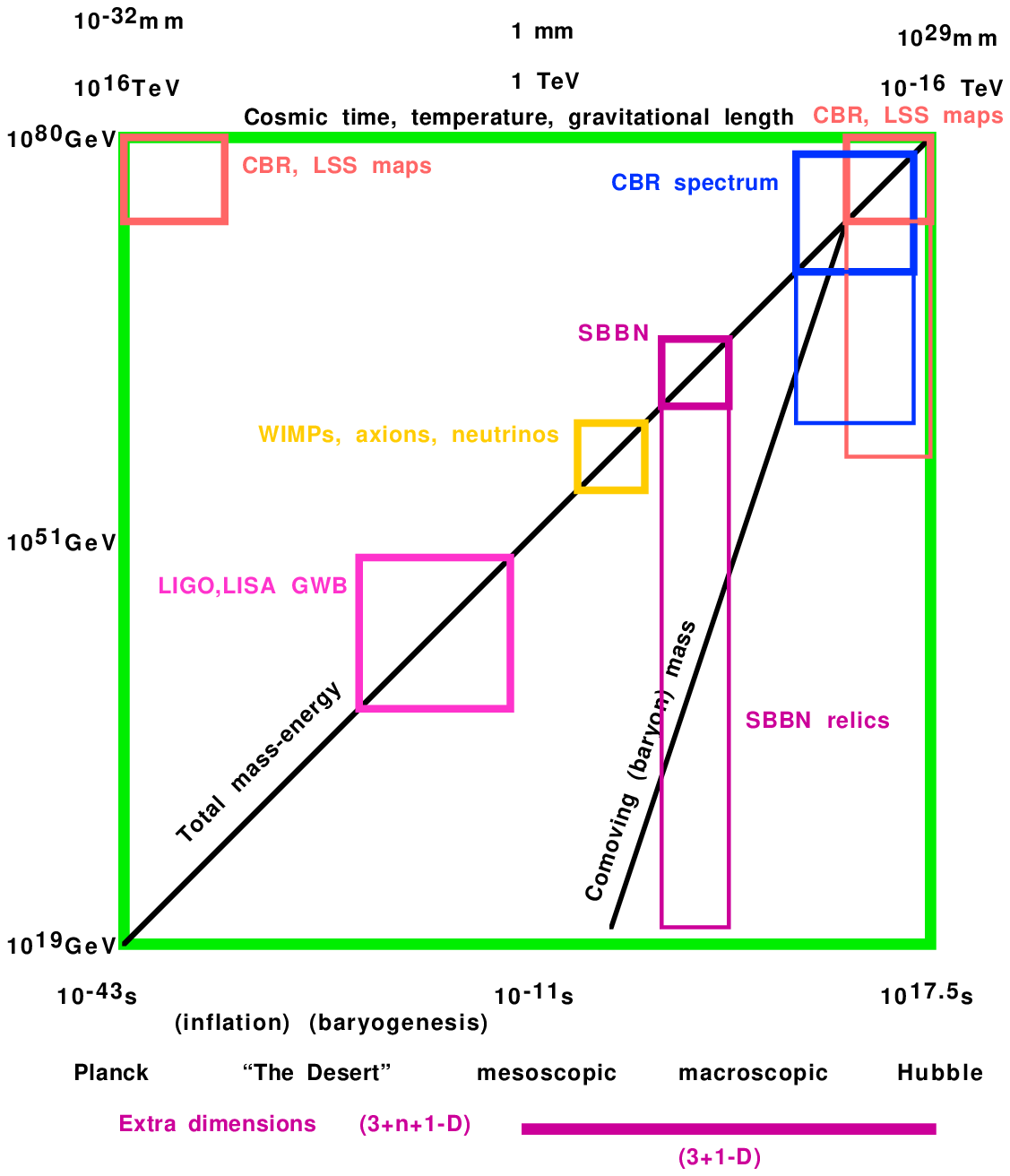,height=5in,width=4in}}
\vspace{0.25in}
\caption{ \label{fig: meso} Summary of evidence about cosmic history, showing scale (in terms of 
total mass-energy) versus cosmic time/temperature. Boxes are
labled by the technique used to constrain events in each domain of time and scale, including
the microwave background anisotropy and spectrum, cosmological nucleosynthesis, and
dark matter production.  The box labeled
LIGO,LISA GWB 
shows the hitherto unexplored region of mesoscopic phenomena which will either  be opened up
or constrained by
gravitational wave astronomy--- the universe earlier than 1 TeV,  when it
may have had more than three spatial dimensions. }
\end{figure}

\end{document}